\documentclass[12pt,thmsa,onecolumn,ukenglish]{article}
\usepackage{amsmath}
\usepackage{sw20lart}

\input{tcilatex}
\begin{document}

\title{Solution of Time-Fractional Korteweg--de Vries Equation in Warm Plasma%
}
\date{}
\author{El-Said A. El-Wakil, Essam M. Abulwafa, \and Emad K. El-shewy and
Abeer A. Mahmoud \\
Theoretical Physics Research Group, Physics Department, \\
Faculty of Science, Mansoura University, Mansoura 35516, Egypt}
\maketitle

\begin{abstract}
The reductive perturbation method has been employed to derive the
Korteweg-de Vries (KdV) equation for small but finite amplitude ion-acoustic
waves. The Lagrangian of the time fractional KdV equation is used in similar
form to the Lagrangian of the regular KdV equation. The variation of the
functional of this Lagrangian leads to the Euler-Lagrange equation that
leads to the time fractional KdV equation. The Riemann-Liouvulle definition
of the fractional derivative is used to describe the time fractional
operator in the fractional KdV equation. The variational-iteration method
given by He is used to solve the derived time fractional KdV equation. The
calculations of the solution with initial condition $A_{0}\sec $h$(cx)^{2}$
are carried out. The result of the present investigation may be applicable
to some plasma environments, such as ionosphere.

\begin{description}
\item $\boldsymbol{Keywords:}$ Ion-acoustic waves; Euler-Lagrange equation,
Riemann-Liouvulle fractional derivative, fractional KdV equation, He's
variational-iteration method.

\item \textbf{PACS:} 05.45.Df, 05.30.Pr.
\end{description}
\end{abstract}

\section{Introduction}

The ion-acoustic solitary wave is one of the fundamental nonlinear wave
phenomena appearing in fluid dynamics [1] and plasma physics [2, 3]. There
are different methods to study nonlinear systems. Washimi and Taniti [4]
were the first to use reductive perturbation method to study the propagation
of a slow modulation of a quasi-monochromatic waves through plasma. And then
the attention has been focused by many authors [5, 6]. The evolution of
small but finite-amplitude solitary waves, studied by means of the
Korteweg--deVries (KdV) equation, is of considerable interest in plasma
dynamics.

Because most classical processes observed in the physical world are
nonconservative, it is important to be able to apply the power of
variational methods to such cases. A method used a Lagrangian that leads to
an Euler-Lagrange equation that is, in some sense, equivalent to the desired
equation of motion. Hamilton's equations are derived from the Lagrangian and
are equivalent to the Euler-Lagrange equation. If a Lagrangian is
constructed using noninteger-order derivatives, then the resulting equation
of motion can be nonconservative. It was shown that such fractional
derivatives in the Lagrangian describe nonconservative forces [7, 8].
Further study of the fractional Euler-Lagrange can be found in the work of
Agrawal [9, 10], Baleanu and coworkers [11, 12] and Tarasov and Zaslavsky
[13, 14]. During the last decades, Fractional Calculus has been applied to
almost every field of science, engineering and mathematics. Some of the
areas where Fractional Calculus has been applied include viscoelasticity and
rheology, electrical engineering, electrochemistry, biology, biophysics and
bioengineering, signal and image processing, mechanics, mechatronics,
physics, and control theory [15].

To the author's knowledge, the problem of time fractional KdV equation in
collisionless plasma has not been addressed in the literature before. So,
our motive here is to study the effects of time fractional parameter on the
electrostatic structures for a system of collisionless plasma consisting of
a mixture of warm ion-fluid and isothermal electrons. We expect that the
inclusion of time fractional parameter will change the properties as well as
the regime of existence of solitons.

Several methods have been used to solve fractional differential equations
such as: the Laplace transform method, the Fourier transform method, the
iteration method and the operational method [16]. Recently, there are some
papers deal with the existence and multiplicity of solution of nonlinear
fractional differential equation by the use of techniques of nonlinear
analysis [17-18]. In this paper, the resultant fractional KdV equation will
be solved using a variational-iteration method (VIM) firstly used by He [19].

This paper is organized as follows: Section 2 is devoted to describe the
formulation of the time-fractional KdV (FKdV) equation using the variational
Euler-Lagrange method. In section 3, variational-Iteration Method 9VIM) is
discussed. The resultant time-FKdV equation is solved approximately using
VIM. Section 5 contains the results of calculations and discussion of these
results.

\section{Basic Equations and KdV Equation}

Consider a collisionless ionization-free unmagnetized plasma consisting of a
mixture of warm ion-fluid and isothermal electrons. The basic equations
describing the system in dimensionless variables is governed by [6]
:\smallskip 
\begin{eqnarray}
\frac{\partial n(x,t)}{\partial t}+\frac{\partial \lbrack n(x,t)u(x,t)]}{%
\partial x} &=&0,  \TCItag{1.a} \\
(\frac{\partial u(x,t)}{\partial t}+u(x,t)\frac{\partial u(x,t)}{\partial x}%
)+\frac{\sigma }{n(x,t)}\frac{\partial p(x,t)}{\partial x}+\frac{\partial
\phi (x,t)}{\partial x} &=&0,  \TCItag{1.b} \\
\lbrack \frac{\partial }{\partial t}+u(x,t)\frac{\partial }{\partial x}%
]p(x,t)+3p(x,t)\frac{\partial u(x,t)}{\partial x} &=&0,  \TCItag{1.c} \\
\frac{\partial ^{2}\phi (x,t)}{\partial x^{2}}+n(x,t)-n_{e}(x,t) &=&0, 
\TCItag{1.d} \\
n_{e}(x,t)-\exp [\phi (x,t)] &=&0.  \TCItag{1.e}
\end{eqnarray}

In the earlier equations $n(x,t)$\ and $n_{e}(x,t)$\ are the densities of
ions and electrons respectively, $u(x,t)$\ is the ion flow velocity, $p(x,t)$%
\ is the ion pressure, $\phi (x,t)$\ is the electric potential, $x$\ is the
space co-ordinate and $t$\ is the time variable. $\sigma =T_{i}/T_{e}<<1$\
is the ratio of the ion temperature $T_{i}$\ to the electron temperature $%
T_{e}$. All these quantities are dimensionless, being normalized in terms of
the following characteristic quantities: $n(x,t)$\ and $n_{e}(x,t)$\ by the
unperturbed electron density $n_{0}$, $u(x,t)$\ by the sound velocity ($%
KT_{e}/m_{i})^{1/2}$\ ; $p(x,t)$ and $\phi (x,t)$\ by $n_{0}KT_{i}$\ and $%
KT_{e}/e$,\ respectively. The time variable $t$\ and the space $x$\ by the
inverse of the plasma frequency $\omega _{pi}^{-1}=(4\pi
e^{2}n_{0}/m_{i})^{-1/2\text{ }}$and the electron Debye length $\lambda
_{D}=(KT_{e}/4\pi e^{2}n_{0})^{1/2}$,\ respectively. $K$\ is the Boltzmann's
constant and $m_{i}$\ is the mass of plasma ion.

According to the general method of reductive perturbation theory, we
introduce the stretched variables

\begin{equation}
\tau =\epsilon ^{\frac{3}{2}}t,\ \xi =\epsilon ^{\frac{1}{2}}(x-vt),  \tag{2}
\end{equation}%
where $v$\ is the unknown phase velocity. All the physical quantities
appeared in (1) are expanded as power series in $\epsilon $\ \ about the
equilibrium values as

\begin{eqnarray}
n(\xi ,\tau ) &=&1+\epsilon n_{1}(\xi ,\tau )+\epsilon ^{2}n_{2}(\xi ,\tau
)+~...\text{,}  \TCItag{3a} \\
u(\xi ,\tau ) &=&\epsilon u_{1}(\xi ,\tau )+\epsilon ^{2}u_{2}(\xi ,\tau
)+~...\text{,}  \TCItag{3b} \\
p(\xi ,\tau ) &=&1+\epsilon \text{ }p_{1}(\xi ,\tau )+\epsilon ^{2}p_{2}(\xi
,\tau )+~...\text{,}  \TCItag{3c} \\
\phi (\xi ,\tau ) &=&\epsilon \phi _{1}(\xi ,\tau )+\epsilon ^{2}\phi
_{2}(\xi ,\tau )+\ ...\text{,}  \TCItag{3d}
\end{eqnarray}%
where $\epsilon $\ represents the amplitude of the perturbation. We impose
the boundary conditions that as $\left\vert \xi \right\vert \rightarrow
\infty ,$\ $n=n_{e}=p=1,$\ $u=\phi =0.\ $Substituting (2) and (3) into the
system of equations (1) and equating coefficients of like powers of $%
\epsilon $. Then, from the lowest, second-order equations in $\epsilon $ and
by elimination the second order perturbed quantities $n_{2}(\xi ,\tau )$, $%
u_{2}(\xi ,\tau )$, $p_{2}(\xi ,\tau )$ and $\phi _{2}(\xi ,\tau )$, we
obtain the following KdV equation for the first-order perturbed potential:

\begin{equation}
\frac{\partial }{\partial \tau }\phi _{1}(\xi ,\tau )+A\text{ }\phi _{1}(\xi
,\tau )\frac{\partial }{\partial \xi }\phi _{1}(\xi ,\tau )+B\text{ }\frac{%
\partial ^{3}}{\partial \xi ^{3}}\phi _{1}(\xi ,\tau )=0,  \tag{4.a}
\end{equation}%
where both the nonlinear coefficient $A$ and the dispersion coefficient $B$
are given by

\begin{equation}
A=\frac{6\sigma +1}{\sqrt{3\sigma +1}}\text{, }B=\frac{1}{2\sqrt{3\sigma +1}}%
\text{.}  \tag{4.b}
\end{equation}

In equation (4.a), $\phi _{1}(\xi ,~\tau )$ is a field variable, $\xi $ is a
space coordinate in the propagation direction of the field and $\tau \in T$ (%
$=[0,T_{0}]$) is the time variable. The resultant KdV equation (4.a) can be
converted into time-fractional KdV equation as follows:

Using a potential function $V(\xi ,~\tau )$ where $\phi _{1}(\xi ,~\tau
)=V_{\xi }(\xi ,~\tau )$ gives the potential equation of the regular KdV
equation (1) in the form

\begin{equation}
V_{\xi \tau }(\xi ,~\tau )+A~V_{\xi }(\xi ,~\tau )v_{\xi \xi }(\xi ,~\tau
)+B~V_{\xi \xi \xi \xi }(\xi ,~\tau )=0\text{,}  \tag{5}
\end{equation}%
where the subscripts denote the partial differentiation of the function with
respect to the parameter. The Lagrangian of this regular KdV equation (4.a)
can be defined using the semi-inverse method [20, 21] as follows.

The functional of the potential equation (5) can be represented by

\begin{eqnarray}
J(V) &=&\dint\limits_{R}d\xi \dint\limits_{T}d\tau \{V(\xi ,\tau
)[c_{1}V_{\xi \tau }(\xi ,\tau )+c_{2}AV_{\xi }(\xi ,\tau )v_{\xi \xi }(\xi
,\tau )  \notag \\
&&+c_{3}BV_{\xi \xi \xi \xi }(\xi ,\tau )]\}\text{,}  \TCItag{6}
\end{eqnarray}%
where $c_{1}$, $c_{2}$ and $c_{3}$ are constants to be determined.
Integrating by parts and taking $V_{\tau }|_{R}=V_{\xi }|_{R}=V_{\xi
}|_{T}=0 $ lead to

\begin{eqnarray}
J(V) &=&\dint\limits_{R}d\xi \dint\limits_{T}d\tau \{V(\xi ,\tau
)[-c_{1}V_{\xi }(\xi ,\tau )V_{\tau }(\xi ,\tau )-\frac{1}{2}c_{2}AV_{\xi
}^{3}(\xi ,\tau )  \notag \\
&&+c_{3}BV_{\xi \xi }^{2}(\xi ,\tau )]\}\text{.}  \TCItag{7}
\end{eqnarray}

The unknown constants $c_{i}$ $(i=1,2,3)$ can be determined by taking the
variation of the functional (7) to make it optimal. Taking the variation of
this functional, integrating each term by parts and make the variation
optimum give the following relation

\begin{equation}
2c_{1}V_{\xi \tau }(\xi ,\tau )+3c_{2}AV_{\xi }(\xi ,\tau )V_{\xi \xi }(\xi
,\tau )+2c_{3}BV_{\xi \xi \xi \xi }(\xi ,\tau )=0\text{.}  \tag{8}
\end{equation}

As this equation must be equal to equation (5), the unknown constants are
given as

\begin{equation}
c_{1}=1/2\text{, }c_{2}=1/3\text{ and }c_{3}=1/2\text{.}  \tag{9}
\end{equation}

Therefore, the functional given by (7) gives the Lagrangian of the regular
KdV equation as

\begin{equation}
L(V_{\tau },~V_{\xi },V_{\xi \xi })=-\frac{1}{2}V_{\xi }(\xi ,\tau )V_{\tau
}(\xi ,\tau )-\frac{1}{6}AV_{\xi }^{3}(\xi ,\tau )+\frac{1}{2}BV_{\xi \xi
}^{2}(\xi ,\tau )\text{.}  \tag{10}
\end{equation}

Similar to this form, the Lagrangian of the time-fractional version of the
KdV equation can be written in the form

\begin{eqnarray}
F(_{0}D_{\tau }^{\alpha }V,~V_{\xi },V_{\xi \xi }) &=&-\frac{1}{2}%
[_{0}D_{\tau }^{\alpha }V(\xi ,\tau )]V_{\xi }(\xi ,\tau )-\frac{1}{6}%
AV_{\xi }^{3}(\xi ,\tau )+\frac{1}{2}BV_{\xi \xi }^{2}(\xi ,\tau )\text{, } 
\notag \\
0 &\leq &\alpha <1\text{,}  \TCItag{11}
\end{eqnarray}%
where the fractional derivative is represented, using the left
Riemann-Liouville fractional derivative definition as [16]

\begin{eqnarray}
_{a}D_{t}^{\alpha }f(t) &=&\frac{1}{\Gamma (k-\alpha )}\frac{d^{k}}{dt^{k}}%
[\int_{a}^{t}d\tau (t-\tau )^{k-\alpha -1}f(\tau )]\text{, }  \notag \\
k-1 &\leq &\alpha \leq 1\text{, }t\in \lbrack a,b]\text{.}  \TCItag{12}
\end{eqnarray}

The functional of the time-FKdV equation can be represented in the form

\begin{equation}
J(V)=\dint\limits_{R}d\xi \dint\limits_{T}d\tau F(_{0}D_{\tau }^{\alpha
}V,~V_{\xi },V_{\xi \xi })\text{,}  \tag{13}
\end{equation}%
where the time-fractional Lagrangian $F(_{0}D_{\tau }^{\alpha }V,~V_{\xi
},V_{\xi \xi })$ is defined by (11).

Following Agrawal's method [9, 10], the variation of functional (13) with
respect to $V(\xi ,\tau )$ leads to

\begin{equation}
\delta J(V)=\dint\limits_{R}d\xi \dint\limits_{T}d\tau \{\frac{\partial F}{%
\partial _{0}D_{\tau }^{\alpha }V}\delta _{0}D_{\tau }^{\alpha }V+\frac{%
\partial F}{\partial V_{\xi }}\delta V_{\xi }+\frac{\partial F}{\partial
V_{\xi \xi }}\delta V_{\xi \xi }\}\text{.}  \tag{14}
\end{equation}

The formula for fractional integration by parts reads [9, 16]

\begin{equation}
\int_{a}^{b}dtf(t)_{a}D_{t}^{\alpha
}g(t)=\int_{a}^{t}dtg(t)_{t}D_{b}^{\alpha }f(t)\text{, \ \ \ }f(t)\text{, }%
g(t)\text{ }\in \lbrack a,~b]\text{,}  \tag{15}
\end{equation}%
where $_{t}D_{b}^{\alpha }$, the right Riemann-Liouville fractional
derivative, is defined by [16]

\begin{eqnarray}
_{t}D_{b}^{\alpha }f(t) &=&\frac{(-1)^{k}}{\Gamma (k-\alpha )}\frac{d^{k}}{%
dt^{k}}[\int_{t}^{b}d\tau (\tau -t)^{k-\alpha -1}f(\tau )]\text{, }  \notag
\\
k-1 &\leq &\alpha \leq 1\text{, }t\in \lbrack a,b]\text{.}  \TCItag{16}
\end{eqnarray}

Integrating the right-hand side of (14) by parts using formula (15) leads to

\begin{equation}
\delta J(V)=\dint\limits_{R}d\xi \dint\limits_{T}d\tau \lbrack _{\tau
}D_{T_{0}}^{\alpha }(\frac{\partial F}{\partial _{0}D_{\tau }^{\alpha }V})-%
\frac{\partial }{\partial \xi }(\frac{\partial F}{\partial V_{\xi }})+\frac{%
\partial ^{2}}{\partial \xi ^{2}}(\frac{\partial F}{\partial V_{\xi \xi }}%
)]\delta V\text{,}  \tag{17}
\end{equation}%
where it is assumed that $\delta V|_{T}=\delta V|_{R}=\delta V_{\xi }|_{R}=0$%
.

Optimizing this variation of the functional $J(V)$, i. e; $\delta J(V)=0$,
gives the Euler-Lagrange equation for the time-FKdV equation in the form

\begin{equation}
_{\tau }D_{T_{0}}^{\alpha }(\frac{\partial F}{\partial _{0}D_{\tau }^{\alpha
}V})-\frac{\partial }{\partial \xi }(\frac{\partial F}{\partial V_{\xi }})+%
\frac{\partial ^{2}}{\partial \xi ^{2}}(\frac{\partial F}{\partial V_{\xi
\xi }})=0\text{.}  \tag{18}
\end{equation}

Substituting the Lagrangian of the time-FKdV equation (11) into this
Euler-Lagrange formula (18) gives

\begin{equation}
-\frac{1}{2}~_{\tau }D_{T_{0}}^{\alpha }V_{\xi }(\xi ,\tau )+\frac{1}{2}%
~_{0}D_{\tau }^{\alpha }V_{\xi }(\xi ,\tau )+AV_{\xi }(\xi ,\tau )V_{\xi \xi
}(\xi ,\tau )+BV_{\xi \xi \xi \xi }(\xi ,\tau )=0\text{.}  \tag{19}
\end{equation}

Substituting for the potential function, $V_{\xi }(\xi ,\tau )=\phi _{1}(\xi
,\tau )=\Phi (\xi ,\tau )$, gives the time-FKdV equation for the state
function $\Phi (\xi ,\tau )$ in the form

\begin{equation}
\frac{1}{2}[_{0}D_{\tau }^{\alpha }\Phi (\xi ,\tau )-_{\tau
}D_{T_{0}}^{\alpha }\Phi (\xi ,\tau )]+A\Phi (\xi ,\tau )\Phi _{\xi }(\xi
,\tau )+B\Phi _{\xi \xi \xi }(\xi ,\tau )=0\text{,}  \tag{20}
\end{equation}%
where the fractional derivatives $_{0}D_{\tau }^{\alpha }$ and $_{\tau
}D_{T_{0}}^{\alpha }$ are, respectively the left and right Riemann-Liouville
fractional derivatives and are defined by (12) and (16).

The time-FKdV equation represented in (20) can be rewritten by the formula

\begin{equation}
\frac{1}{2}~_{0}^{R}D_{\tau }^{\alpha }\Phi (\xi ,\tau )+A~\Phi (\xi ,\tau
)\Phi _{\xi }(\xi ,\tau )+B~\Phi _{\xi \xi \xi }(\xi ,\tau )=0\text{,} 
\tag{21}
\end{equation}%
where the fractional operator $_{0}^{R}D_{\tau }^{\alpha }$ is called Riesz
fractional derivative and can be represented by [10, 16]

\begin{eqnarray}
~_{0}^{R}D_{t}^{\alpha }f(t) &=&\frac{1}{2}[_{0}D_{t}^{\alpha
}f(t)+~(-1)^{k}{}_{t}D_{T_{0}}^{\alpha }f(t)]  \notag \\
&=&\frac{1}{2}\frac{1}{\Gamma (k-\alpha )}\frac{d^{k}}{dt^{k}}%
[\int_{a}^{t}d\tau |t-\tau |^{k-\alpha -1}f(\tau )]\text{, }  \notag \\
k-1 &\leq &\alpha \leq 1\text{, }t\in \lbrack a,b]\text{.}  \TCItag{22}
\end{eqnarray}

The nonlinear fractional differential equations have been solved using
different techniques [16-20]. In this paper, a variational-iteration method
(VIM) [21, 22] has been used to solve the time-FKdV equation that formulated
using Euler-Lagrange variational technique.

\section{Variational-Iteration Method}

Variational-iteration method (VIM) [21] has been used successfully to solve
different types of integer nonlinear differential equations [22, 23]. Also,
VIM is used to solve linear and nonlinear fractional differential equations
[24, 25]. This VIM has been used in this paper to solve the formulated
time-FKdV equation.

A general Lagrange multiplier method is constructed to solve non-linear
problems, which was first proposed to solve problems in quantum mechanics
[21]. The VIM is a modification of this Lagrange multiplier method [22]. The
basic features of the VIM are as follows. The solution of a linear
mathematical problem or the initial (boundary) condition of the nonlinear
problem is used as initial approximation or trail function. A more highly
precise approximation can be obtained using iteration correction functional.

Considering a nonlinear partial differential equation consists of a linear
part $\overset{\symbol{94}}{L}U(x,t)$, nonlinear part $\overset{\symbol{94}}{%
N}U(x,t)$ and a free term $f(x,t)$ represented as

\begin{equation}
\overset{\symbol{94}}{L}U(x,t)+\overset{\symbol{94}}{N}U(x,t)=f(x,t)\text{,}
\tag{23}
\end{equation}%
where $\overset{\symbol{94}}{L}$ is the linear operator and $\overset{%
\symbol{94}}{N}$ is the nonlinear operator. According to the VIM, the ($n+1$)%
\underline{th} approximation solution of (23) can be given by the iteration
correction functional as [21, 22]

\begin{eqnarray}
U_{n+1}(x,t) &=&U_{n}(x,t)+\int_{0}^{t}d\tau \lambda (\tau )[\overset{%
\symbol{94}}{L}U_{n}(x,\tau )+\overset{\symbol{94}}{N}\hat{U}_{n}(x,\tau
)-f(x,\tau )]\text{, }  \notag \\
n &\geq &0\text{,}  \TCItag{24}
\end{eqnarray}%
where $\lambda (\tau )$ is a Lagrangian multiplier and $\hat{U}_{n}(x,\tau )$
is considered as a restricted variation function, i. e; $\delta \hat{U}%
_{n}(x,\tau )=0$. Extreme the variation of the correction functional (24)
leads to the Lagrangian multiplier $\lambda (\tau )$. The initial iteration
can be used as the solution of the linear part of (23) or the initial value $%
U(x,0)$. As n tends to infinity, the iteration leads to the exact solution
of (23), i. e;

\begin{equation}
U(x,t)=\underset{n\rightarrow \infty }{\lim }U_{n}(x,t)\text{.}  \tag{25}
\end{equation}%
\qquad For linear problems, the exact solution can be given using this
method in only one step where its Lagrangian multiplier can be exactly
identified.

\section{Solution of the time-FKdV equation}

The time-FKdV equation represented by (21) can be solved using the VIM by
the iteration correction functional (24) as follows:

Affecting from left by the fractional operator on (21) leads to

\begin{eqnarray}
\frac{\partial }{\partial \tau }\Phi (\xi ,\tau ) &=&~_{0}^{R}D_{\tau }^{\
\alpha -1}\Phi (\xi ,\tau )|_{\tau =0}\frac{\tau ^{\alpha -2}}{\Gamma
(\alpha -1)}  \notag \\
&&-\ _{0}^{R}D_{\tau }^{\ 1-\alpha }[A~\Phi (\xi ,\tau )\frac{\partial }{%
\partial \xi }\Phi (\xi ,\tau )+B~\frac{\partial ^{3}}{\partial \xi ^{3}}%
\Phi (\xi ,\tau )]\text{, }  \notag \\
0 &\leq &\alpha \leq 1\text{, }\tau \in \lbrack 0,T_{0}]\text{,}  \TCItag{26}
\end{eqnarray}%
where the following fractional derivative property is used [16]

\begin{eqnarray}
\ _{a}^{R}D_{b}^{\ \alpha }[\ _{a}^{R}D_{b}^{\ \beta }f(t)]
&=&~_{a}^{R}D_{b}^{\ \alpha +\beta }f(t)-\overset{k}{\underset{j=1}{\sum }}\
_{a}^{R}D_{b}^{\ \beta -j}f(t)|_{t=a}~\frac{(t-a)^{-\alpha -j}}{\Gamma
(1-\alpha -j)}\text{, }  \notag \\
k-1 &\leq &\beta <k\text{.}  \TCItag{27}
\end{eqnarray}

As $\alpha <1$, the Riesz fractional derivative $_{0}^{R}D_{\tau }^{\ \alpha
-1}$ is considered as Riesz fractional integral $_{0}^{R}I_{\tau }^{1-\alpha
}$ that is defined by [10, 16]

\begin{eqnarray}
\ _{0}^{R}I_{\tau }^{\ \alpha }f(t) &=&\frac{1}{2}[_{0}I_{\tau }^{\ \alpha
}f(t)\ +~_{\tau }I_{b}^{\ \alpha }f(t)]  \notag \\
&=&\frac{1}{2}\frac{1}{\Gamma (\alpha )}\int_{a}^{b}d\tau |t-\tau |^{\alpha
-1}f(\tau )\text{, }\alpha >0\text{,}  \TCItag{28}
\end{eqnarray}%
where $_{0}I_{\tau }^{\ \alpha }f(t)$\ and $_{\tau }I_{b}^{\ \alpha }f(t)$
are the left and right Riemann-Liouvulle fractional integrals, respectively
[16].

The iterative correction functional of equation (26) is given as

\begin{eqnarray}
\Phi _{n+1}(\xi ,\tau ) &=&\Phi _{n}(\xi ,\tau )+\int_{0}^{\tau }d\tau
^{\prime }\lambda (\tau ^{\prime })\{\frac{\partial }{\partial \tau ^{\prime
}}\Phi _{n}(\xi ,\tau ^{\prime })  \notag \\
&&-~_{0}^{R}I_{\tau ^{\prime }}^{1-\alpha }\Phi _{n}(\xi ,\tau ^{\prime
})|_{\tau ^{\prime }=0}\frac{\tau ^{\prime \alpha -2}}{\Gamma (\alpha -1)} 
\notag \\
&&+\ _{0}^{R}D_{\tau ^{\prime }}^{\ 1-\alpha }[A~\overset{\symbol{126}}{\Phi 
}_{n}(\xi ,\tau ^{\prime })\frac{\partial }{\partial \xi }\overset{\symbol{%
126}}{\Phi }_{n}(\xi ,\tau ^{\prime })+B\frac{\partial ^{3}}{\partial \xi
^{3}}\overset{\symbol{126}}{\Phi }_{n}(\xi ,\tau ^{\prime })]\}\text{,} 
\TCItag{29}
\end{eqnarray}%
where $n\geq 0$ and the function $\overset{\symbol{126}}{\Phi }_{n}(\xi
,\tau )$ is considered as a restricted variation function, i. e; $\delta 
\overset{\symbol{126}}{\Phi _{n}}(\xi ,\tau )=0$. The extreme of the
variation of (29) using the restricted variation function leads to

\begin{eqnarray*}
\delta \Phi _{n+1}(\xi ,\tau ) &=&\delta \Phi _{n}(\xi ,\tau
)+\int_{0}^{\tau }d\tau ^{\prime }\lambda (\tau ^{\prime })~\delta \frac{%
\partial }{\partial \tau ^{\prime }}\Phi _{n}(\xi ,\tau ^{\prime }) \\
&=&\delta \Phi _{n}(\xi ,\tau )+\lambda (\tau )~\delta \Phi _{n}(\xi ,\tau
)-\int_{0}^{\tau }d\tau ^{\prime }\frac{\partial }{\partial \tau ^{\prime }}%
\lambda (\tau ^{\prime })~\delta \Phi _{n}(\xi ,\tau ^{\prime })=0\text{.}
\end{eqnarray*}

This relation leads to the stationary conditions $1+\lambda (\tau )=0$ and $%
\frac{\partial }{\partial \tau ^{\prime }}\lambda (\tau ^{\prime })=0$,
which leads to the Lagrangian multiplier as $\lambda (\tau ^{\prime })=-1$.
\ Therefore, the correction functional (29) is given by the form

\begin{eqnarray}
\Phi _{n+1}(\xi ,\tau ) &=&\Phi _{n}(\xi ,\tau )-\int_{0}^{\tau }d\tau
^{\prime }\{\frac{\partial }{\partial \tau ^{\prime }}\Phi _{n}(\xi ,\tau
^{\prime })  \notag \\
&&-~_{0}^{R}I_{\tau ^{\prime }}^{1-\alpha }\Phi _{n}(\xi ,\tau ^{\prime
})|_{\tau ^{\prime }=0}\frac{\tau ^{\prime \alpha -2}}{\Gamma (\alpha -1)} 
\notag \\
&&+\ _{0}^{R}D_{\tau ^{\prime }}^{\ 1-\alpha }[A~\Phi _{n}(\xi ,\tau
^{\prime })\frac{\partial }{\partial \xi }\Phi _{n}(\xi ,\tau ^{\prime })+B~%
\frac{\partial ^{3}}{\partial \xi ^{3}}\Phi _{n}(\xi ,\tau ^{\prime })]\}%
\text{,}  \TCItag{30}
\end{eqnarray}%
where  $n\geq 0$.

In Physics, if $\tau $ denotes the time-variable, the right
Riemann-Liouville fractional derivative is interpreted as a future state of
the process. For this reason, the right-derivative is usually neglected in
applications, when the present state of the process does not depend on the
results of the future development [9]. Therefore, the right-derivative is
used equal to zero in the following calculations.

The zero order correction of the solution can be taken as the initial value
of the state variable, which is taken in this case as

\begin{equation}
\Phi _{0}(\xi ,\tau )=\Phi (\xi ,0)=A_{0}\sec \text{h}^{2}(c\xi )\text{.} 
\tag{31}
\end{equation}

where $A_{0}=\frac{3v}{A}$ and $c=\frac{1}{2}\sqrt{\frac{v}{B}}$ are
constants.

Substituting this zero order approximation into (30) and using the
definition of the fractional derivative (22) lead to the first order
approximation as

\begin{eqnarray}
\Phi _{1}(\xi ,\tau ) &=&A_{0}\sec \text{h}^{2}(c\xi )  \notag \\
&&+2A_{0}c~\sinh (c\xi )~\sec \text{h}^{3}(c\xi )  \notag \\
&&\ast \lbrack 4c^{2}B+(A_{0}A-12c^{2}B)\sec h^{2}(c\xi )]\frac{\tau
^{\alpha }}{\Gamma (\alpha +1)}\text{.}  \TCItag{32}
\end{eqnarray}

Substituting this equation into (30), using the definition (22) and the
Maple package lead to the second order approximation in the form

\begin{eqnarray}
\Phi _{2}(\xi ,\tau ) &=&A_{0}\sec \text{h}^{2}(c\xi )  \notag \\
&&+2A_{0}c~\sinh (c\xi )~\sec \text{h}(c\xi )^{3}  \notag \\
&&\ast \lbrack 4c^{2}B+(A_{0}A-12c^{2}B)\sec \text{h}(c\xi )^{2}]\frac{\tau
^{\alpha }}{\Gamma (\alpha +1)}  \notag \\
&&+2A_{0}c^{2}\sec \text{h}(c\xi )^{2}  \notag \\
&&\ast \lbrack 32c^{4}B^{2}+16c^{2}B(5A_{0}A-63c^{2}B)\sec \text{h}(c\xi
)^{2}  \notag \\
&&+2(3A_{0}^{2}A^{2}-176A_{0}c^{2}AB+1680c^{4}B^{2})\sec \text{h}(c\xi )^{4}
\notag \\
&&-7(A_{0}^{2}A^{2}-42A_{0}c^{2}AB+360c^{4}B^{2})\sec \text{h}(c\xi )^{6}]%
\frac{\tau ^{2\alpha }}{\Gamma (2\alpha +1)}  \notag \\
&&+4A_{0}^{2}c^{3}\sinh (c\xi )~\sec \text{h}(c\xi )^{5}  \notag \\
&&\ast \lbrack 32c^{4}B^{2}+24c^{2}B(A_{0}A-14c^{2}B)\sec \text{h}(c\xi )^{2}
\notag \\
&&+4(A_{0}^{2}A^{2}-32A_{0}c^{2}AB+240c^{4}B^{2})\sec \text{h}(c\xi )^{4} 
\notag \\
&&-5(A_{0}^{2}A^{2}-24A_{0}c^{2}AB+144c^{4}B^{2})\sec \text{h}(c\xi )^{6}] 
\notag \\
&&\ast \frac{\Gamma (2\alpha +1)}{[\Gamma (\alpha +1)]^{2}}\frac{\tau
^{3\alpha }}{\Gamma (3\alpha +1)}  \TCItag{33}
\end{eqnarray}

The higher order approximations can be calculated using the Maple or the
Mathematica package to the appropriate order where the infinite
approximation leads to the exact solution.

\section{Results and calculations}

For a small amplitude ion-acoustic solitary wave in unmagnetized
collisionless plasma consisting of a mixture of warm ion-fluid and
isothermal electrons, we have obtained the Korteweg-de Vries equation by
using the reductive perturbation method. The time fractional Korteweg-de
Vries equation has been derived using the variational technique [9, 20]. The
Riemann-Liouvulle fractional derivative is used to describe the time
fractional operator in the time-FKdV equation. He's variational-iteration
method [21, 22] is used to solve the derived time-FKdV equation. The
calculations in this work is carried out using the fourth order
approximation of the VIM.

However, since one of our motivations was to study effects of the ratio of
the ion temperature to the electron temperature $\sigma $ and the time
fractional order $\alpha $ on the existence of solitary waves. The present
system admits a solitary wave solution for any order approximation. In Fig
(1), a profile of the solution of the time FKdV equation as a solitary pulse
is obtained. Figure (2) shows that the both the amplitude and the width of
the solitary wave decrease with the increase of the temperatures ratio $%
\sigma $. Also, the time fractional order $\alpha $ and the velocity $v$
increase the soliton amplitude as shown in Fig (3). In summery, it has been
found that amplitude of the ion-acoustic waves as well as parametric regime
where the solitons can exist is sensitive to the temperatures ratio $\sigma $
and the time fractional order $\alpha $. Moreover, the time fractional order 
$\alpha $ plays a role of higher order perturbation theory in increasing the
soliton amplitude. The application of our model might be particularly
interesting in some plasma environments, such as ionosphere.\pagebreak

\section{Figure Captions}

Fig. (1): The electrostatic potential $\Phi (\xi ,~\tau )$ against the
position $\xi $ and the time $\tau $ for $\sigma =0.5$, $v=0.4$ and $\alpha
=0.5$.

Fig. (2): The electrostatic potential $\Phi (\xi ,~\tau )$ against the
position $\xi $ and the temperatures ratio $\sigma $ at $\tau =1$, $v=0.4$
and $\alpha =0.5$.

Fig. (3): The electrostatic potential $\Phi (\xi ,~\tau )$ against the
velocity $v$ and the fractional order $\alpha $ at $\xi =0$, $\tau =1$ and $%
\sigma =0.5$.\pagebreak 

\end{document}